# Exponentially selective molecular sieving through angstrom pores

P. Z. Sun[1,2], M. Yagmurcukardes[3,4,5], R. Zhang[1], W. J. Kuang[1], M. Lozada-Hidalgo[1], F. C. Wang[6], F. M. Peeters[3,4], I. V. Grigorieva[1], A. K. Geim[1,2]

[1]Department of Physics and Astronomy, University of Manchester, Manchester M13 9PL, UK
[2]National Graphene Institute, University of Manchester, Manchester M13 9PL, UK
[3]Department of Physics, University of Antwerp, Groenenborgerlaan 171, B-2020 Antwerp, Belgium
[4]NANOlab Center of Excellence, Groenenborgerlaan 171, B-2020 Antwerp, Belgium
[5]Department of Photonics, Izmir Institute of Technology, 35430 Izmir, Turkey
[6]Chinese Academy of Sciences Key Laboratory of Mechanical Behavior and Design of Materials, Department of Modern Mechanics, University of Science and Technology of China, Hefei, Anhui 230027, China

*Two-dimensional crystals with angstrom-scale pores are widely considered as candidates for a next generation of molecular separation technologies aiming to provide extreme selectivity combined with high flow rates. Here we study gas transport through individual graphene pores with an effective diameter of ~2 angstroms, or about one missing carbon ring, which are created reproducibly by a short-time exposure to a low-kV electron beam. Helium and hydrogen permeate easily through these pores whereas larger molecules such as xenon and methane are blocked. Permeating gases experience activation barriers that increase quadratically with the kinetic diameter, and the transport process crucially involves surface adsorption. Our results reveal underlying mechanisms for the long sought-after exponential selectivity and suggest the bounds on possible performance of porous two-dimensional membranes.*

Two-dimensional (2D) membranes with a high density of angstrom-scale pores can be made by engineering defects in 2D crystals[1-9] or, perhaps more realistically in terms of applications, by growing intrinsically porous crystals such as, e.g., graphynes[10-12]. Interest in angstroporous 2D materials is strongly stimulated by potential applications, particularly for gas separation as an alternative to polymeric membranes employed by industry[3,13]. On the one hand, the atomic thickness of 2D materials implies a relatively high permeability as compared to traditional 3D membranes. On the other hand, angstrom-scale pores with effective sizes $d_P$ smaller than the kinetic diameter $d_K$ of molecules should pose substantial barriers for their translocation, which is predicted to result in colossal selectivities $S > 10^{10}$, even for gases with fractionally (~25%) different $d_K$ such as, for example, $H_2$ and $CH_4$[1,14,15]. This unique combination of material properties holds a promise of better selectivity-permeability tradeoffs than those possible by conventional membranes[3,13]. At present, this optimistic assessment is based mostly on theoretical modeling. Experimental clarity has so far been achieved only for the classical regime of $d_P > d_K$ where the flow is governed by the Knudsen equation, and the resulting selectivities arise from modest differences in thermal velocities of gases having different molecular masses $m$[7-9,16]. For smaller pores with $d_P \approx d_K$, $S$ up to 10-100 have been reported for monolayer graphene[5,8], and even higher selectivities (~$10^4$) were achieved for pores with an estimated diameter of ~3.5 Å in bilayer graphene[4]. Still, this is many orders of magnitude smaller than $S$ predicted for the activated transport regime $d_P < d_K$[1,14,15]. Little remains known about the latter regime that is proved difficult so far to probe experimentally[5,8,9]. The lack of understanding is further exacerbated by prohibitive computational costs of simulating molecular permeation in the activated regime[17-20].

We were able to achieve and study this regime by using an alternative approach of creating pores in graphene, which involved low-energy electron irradiation in a scanning electron microscope. Our experimental devices were micrometer-size cavities sealed with monolayer graphene (Fig. 1a). The microcavities were fabricated from graphite monocrystals, using lithography and dry etching, and had internal diameters of 1 to 3 μm and depth of ~100 nm (Supplementary Information). Large exfoliated graphene crystals were then transferred in air on top of the microcavities, creating 'atomically-tight' sealing[21]. The sealing was tested by placing the devices into a He atmosphere and monitoring changes in graphene membrane's position by atomic force microscopy (AFM) (Fig. 1b). We selected only the devices that were completely impermeable to He (see Supplementary Information and ref. 21). Next, the membranes were subjected to a ≤10 keV electron beam for a few seconds (doses of 0.1-0.2 μC cm$^{-2}$), and the devices were He-leak tested post-exposure. The procedure was repeated, until a leak appeared indicating a damage induced by electrons (Fig. 1c). In most cases (>95%), only a single pore was created. This was proven by sealing the leak with sparsely dispersed Au nanoparticles (see Supplementary Information and Fig. S2), following the approach suggested in ref. 5. Furthermore, after such a pore appeared, we could increase the radiation dose by 100 times but normally were unable to create additional pores or even modify the existing ones (Fig. S3). We attribute formation of the pores to chemical etching of graphene with locally adsorbed water activated by the electron beam[22-24] and speculate that further exposure had covered graphene with cross-linked hydrocarbons[25] which protected it from continuous water-mediated damage. Unfortunately, no microscopy technique can visualize the pores' atomic structure, especially because they are extremely sparse and graphene is covered with hydrocarbon contamination[4,5,26] (see Supplementary Information).

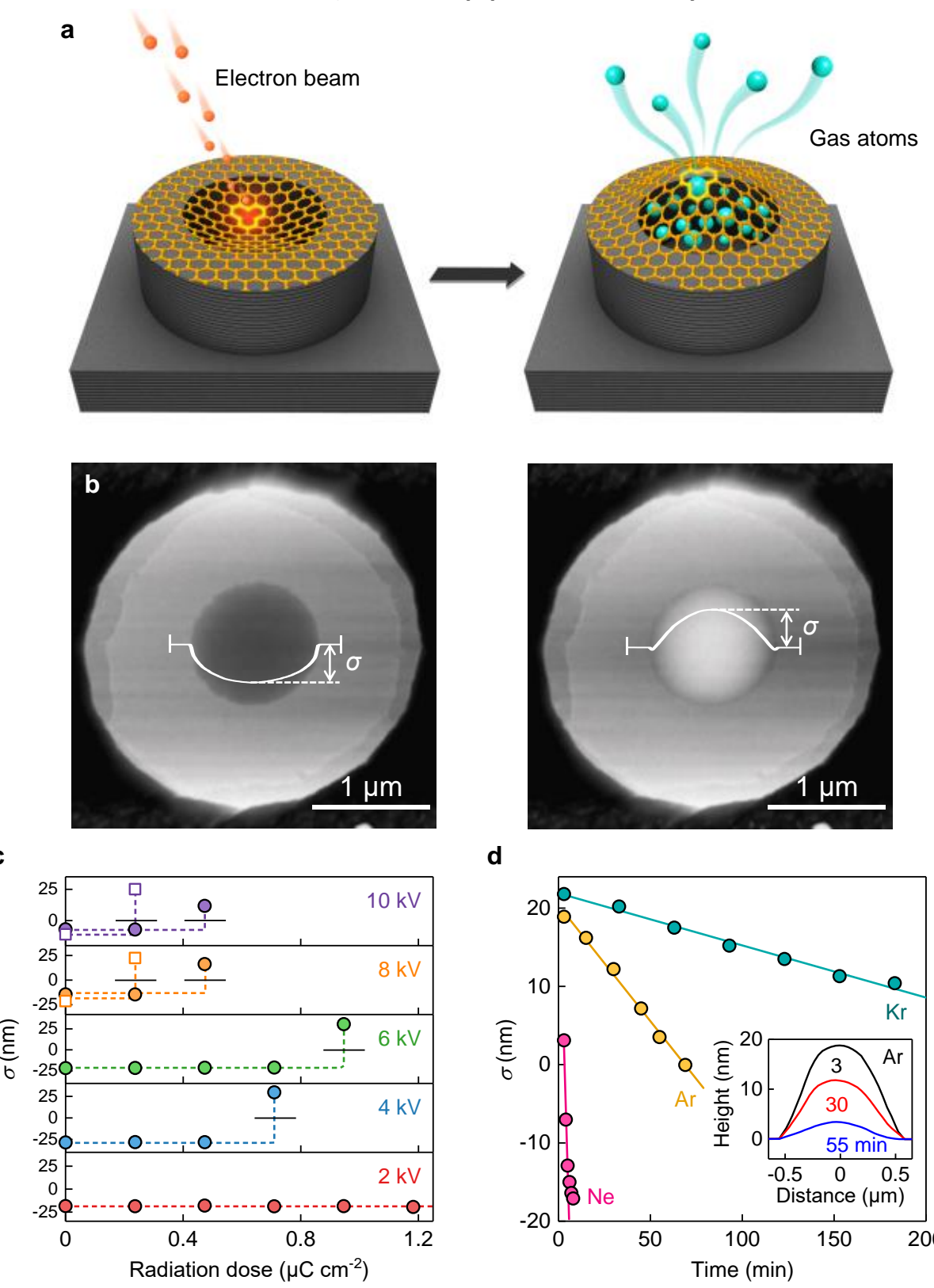

***Fig. 1 | Creating defects in suspended graphene. a**, Schematic of our devices. Left: Monolayer graphene sealing a microcavity was bombarded with electrons. Initially, the membrane sagged inside the cavity due to adhesion to the side walls[4,5,21]. Right: After pressurization, defected membranes bulged out. **b**, AFM images of the same device before (left) and after (right) its exposure to 10 keV electrons; dose of 0.5 μC cm$^{-2}$. Both images were taken after storing the device in Kr at 3 bar for 10 days. The white curves are height profiles along the membrane diameter[21]. σ is the membrane's central position measured with respect to graphite's top surface. The grey scale is given by σ ≈–15 and +24 nm in the left and right images, respectively. **c**, Examples of σ as a function of radiation dose and acceleration voltage. Each point is taken after pressurizing the devices in 3-bar Kr. Dashed lines: guides to the eye; short black lines: σ = 0. **d**, σ(t) for a device with the medium-size pore denoted as type 2, after pressurizing it with various gases (color coded). Solid curves: best linear fits. Inset: representative height profiles for a deflating device with Ar inside.*

The defected membranes prepared as described above were subjected to further permeation tests using various gases (namely, He, Ne, Ar, Kr, Xe, $H_2$, $CO_2$, $O_2$, $N_2$ and $CH_4$). To this end, the devices were placed in a chamber containing a mixture of air at 1 bar (to match the air captured inside during fabrication) and the tested gas at a partial pressure $P$ of typically ≥ 3 bar. Storage for 2-20 days, depending on the gas, allowed pressures inside and outside to equalize so that the membranes reached stable-in-time positions. After taking the devices back into air, graphene membranes would normally bulge out (Fig. 1a,b) and then gradually deflate, which was monitored by AFM (Fig. 1d). For quantitative analysis, we recorded the central position $\sigma$ of bulged membranes (Fig. 1b) as a function of time $t$. Initially, $\sigma$ evolved linearly with $t$, indicating a constant outflow of the tested gas (Fig. 1d), until its partial pressure inside dropped leading to saturation in $\sigma(t)$, in agreement with refs. 4,5. We used the initial slope to evaluate the permeation rate $\Gamma$ for each gas, as described in Supplementary Information. Repeating this procedure at different $P$, we confirmed that $\Gamma \propto P$ (Fig. S1) and, therefore, the pores could be characterized by their $P$-independent permeance $\Gamma^* = \Gamma/P$. For slowly permeating gases, our range of $\Gamma^*$ was limited by observational times of several days, which yielded a permeance of ~$10^{-31}$ mol s$^{-1}$ Pa$^{-1}$, or less than one atom per minute escaping the cavity. It is due to this exceptional sensitivity that we could detect flows through individual pores in the activated transport regime, which would be difficult if not impossible to access otherwise[4,5,9,21]. As for the upper limit on $\Gamma^*$, it was determined by the required time of ~3 min to obtain an AFM image after taking devices from the gas chamber, which translates into ~$10^{-23}$ mol s$^{-1}$ Pa$^{-1}$, if using high $P$ = 10 bar and our largest cavities.

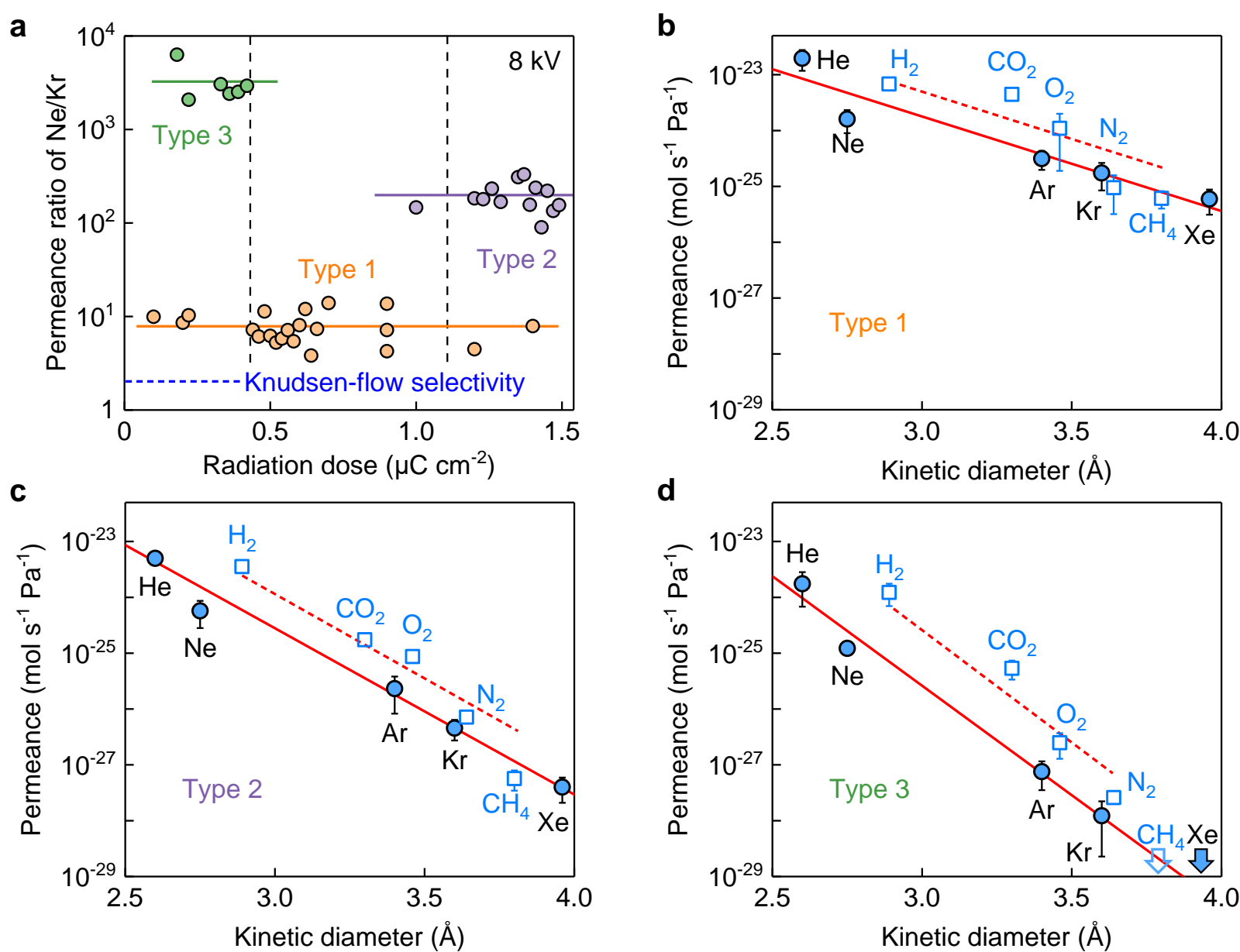

***Fig. 2| Gas selectivity for graphene pores created by electron bombardment. a**, Selectivity between Ne and Kr as a function of the dosage at which the pores appeared under an 8-kV electron beam. Each symbol denotes a different device. Three tight groups are emphasized by their color with the solid lines indicating the average S for each group. Vertical lines: guides to the eye indicating the threshold doses for different pore types. **b-d**, $\Gamma^*$ for the 3 types of pores using 10 different gases, as annotated in the panels. Error bars: SD for typically 6 but minimum 3 devices. Solid curves in (**b-d**): best fits to $\Gamma^* \propto \exp(-\alpha d_K)$ for noble gases with α being constants. Because of the limited range of $d_K$, the data fit equally well with $\Gamma^* \propto \exp(-\alpha d_K^2)$ (not shown). Dashed curves: guides to the eye for diatomic gases. The arrows in (**d**) refer to undetectable permeation for Xe and $CH_4$.*

Our measurements of $\Gamma^*$ are summarized in Fig. 2 using more than 40 devices with each one used to probe several gases. Only 3 distinct types of pores were observed. This is illustrated by Fig. 2a that compares $\Gamma^*$ for Ne and Kr (30% different $d_K$). The measured selectivities $S = \Gamma^*(\mathrm{Ne})/\Gamma^*(\mathrm{Kr})$ fall into 3 clearly separated groups. Small scattering around the average $S$ within each group can be attributed to random local strain or curvature[21]. We refer to the groups as type 1, 2 and 3 pores, according to their $S$. Using other voltages between 4 and 10 kV, only the same three types of pores were observed (Supplementary Information). Figure 2a also shows that the radiation dose at which a pore appeared can serve as a good predictor of its type, with low and high doses favoring type 3 and 2 pores, respectively.

Characteristics of each pore type are detailed in Fig. 2b-d. All the pores exhibited exponential dependences $\Gamma^*(d_K)$ with type 3 being most selective, followed by type 2 and 1. Judging by their permeance, type-1 pores are similar to those created by ultraviolet-induced oxidation[5]. Within our sensitivity limits, the smallest (type-3) pores were completely impermeable to Xe and $CH_4$ yielding selectivity $>10^7$ with respect to He or $H_2$, which is higher than $S$ for any type of membranes reported in the literature. Surprisingly, diatomic gases exhibited systematically higher $\Gamma^*$ than noble gases (Fig. 2). This cannot be due to the elongated shape of diatomic molecules because $d_K$ corresponds to the smallest cross-section[27], that is, the most favorable orientation for translocation. Figure 2 also shows that the observed permeation was controlled mainly by spatial confinement rather than, e.g., chemical affinity: otherwise, translocation of molecules containing certain atoms would fall out of the monotonic sequences.

To investigate the involved sieving mechanisms, we measured temperature ($T$) dependences of $\Gamma^*$ for all pore types. An example of such measurements is shown in Fig. 3a whereas Fig. 3b plots the extracted activation energies $E_A$, using $\Gamma^* = \nu \exp(-E_A/k_B T)$ where $k_B$ is the Boltzmann constant and $\nu$ is the attempt rate. If plotted as a function of $d_K^2$ (rather than $d_K$) our data closely follow $E_A = \alpha(d_K^2 - d_0^2)$. This allows the following interpretation. The pores have an empty space with the diameter $d_0$ which is free from graphene's electron clouds (inset of Fig. 3b). To 'squeeze' through the pore, molecules must disturb a region of approximately $\pi(d_K^2 - d_0^2)/4$ in size, and both electronic and elastic contributions are expected to scale with this value (Fig. S4). The same $\alpha$ for all three pore types strongly supports this interpretation, indicating that $E_A$ is determined by the graphene properties, independent of pores' configurations and diameters.

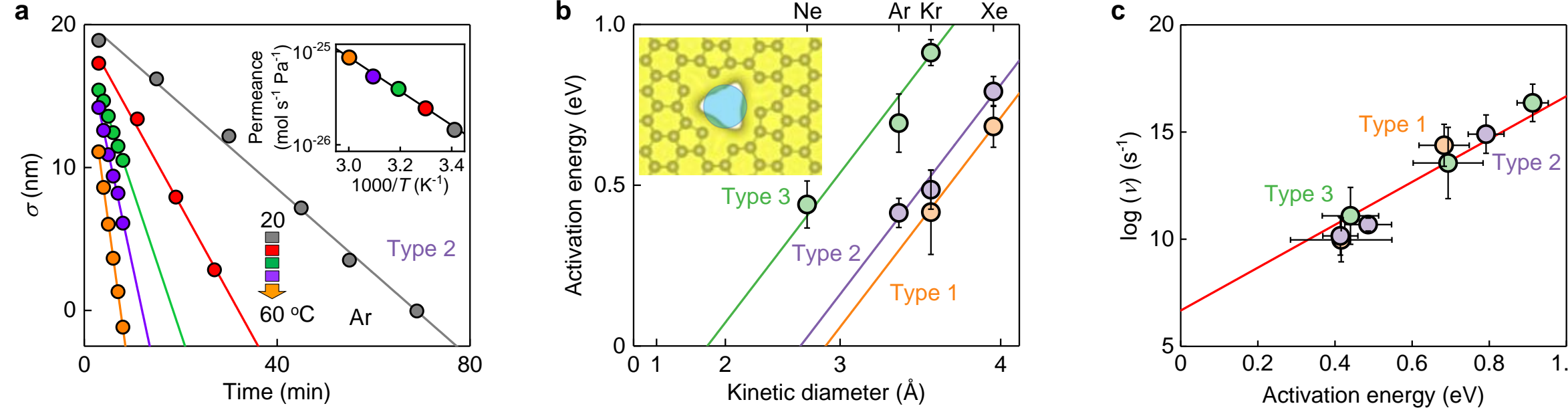

***Fig. 3| Characterizing the angstrom pores. a**, Example of the measured T dependences for type-2 pores (color coded T). Symbols: experimental data for Ar. Solid lines: linear fits. Inset: resulting Arrhenius plot (same color-coding). Solid curve: best fit yielding $E_A \approx 0.4$ eV. **b**, $E_A$ for noble gases and different pore types shown as a function of $d_K$ (note the nonlinear x axis). Symbols: experimental data with error bars showing SD. Solid curves: best fits with $E_A = \alpha (d_K^2 - d_0^2)$ using same α. Inset: possible atomic-scale defect (Supplementary Information) with $d_0$ close to that of type-2 pores (blue circle's diameter is 2.5 Å). **c**, Attempt rates ν at 1 bar for the same gases and $E_A$ as in (**b**). The solid line: best fit using $1/\beta$ = 40 meV[28,29].*

Next, we analyze the pre-exponential factors $\nu$ (Fig. 3c), which were also found from the measured $T$ dependences such as in Fig. 3a. For atoms arriving from the bulk, their attempt rate is given by $\nu_0 = AP/(2\pi m k_B T)^{1/2}$ where $A$ is the effective pore area[5,17,18,21], which yields $\nu_0$ of the order of $10^8$ $s^{-1}$ at 1 bar for all our pores and gases. In contrast, the experiment yielded many orders of magnitude higher $\nu$ (Fig. 3c). This unambiguously indicates that translocating atoms come not from the bulk but mostly through adsorption and surface diffusion[17,18,20]. This has the following consequences. First, the actual translocation barriers should be somewhat higher than the measured $E_A$, as the latter values are reduced by the adsorption energy[8,18] (see Supplementary Information). Second, the adsorption-diffusion mechanism should favor permeation of stronger-adsorbed diatomic gases, in conceptual agreement with their systematically higher $\Gamma^*$ compared to noble gases (Fig. 2).

In the limit of zero $E_A$, the attempt rate in Fig. 3c extrapolates close to $\nu_0$, as expected because this limit corresponds to the Knudsen flow. On the other hand, the strong dependence $\nu \approx \nu_0 \exp(\beta E_A)$ in Fig. 3c is surprising. We speculate that it can be due to entropy loss during the surface-transport permeation process, as discussed in the literature[28,29], and is a result of an increasingly large area that supplies gas molecules to the pore mouth, which rapidly grows with increasing $E^2$[2,28] (see Supplementary Information). Note that polymeric membranes exhibit similar $\nu(E_A)$ dependences with a universal, material-independent coefficient $\beta \approx 1/(40 \text{ meV})$[28,29] which also matches well our results (Fig. 3c). The origins of such universality remain unknown[28,29].

To conclude, our work provides a feedback for extensive theoretical studies of molecular transport through angstrom-scale pores and reveals puzzling features of the activated-transport mechanism. The mechanism critically involves adsorption and surface diffusion, which places strong constraints on the pore sizes required to reach high selectivity. For example, the found dependence $\nu \approx \nu_0 \exp(\beta E_A)$ counteracts the Arrhenius behavior $\exp(-E_A/k_BT)$ and strongly reduces $S$ for any given pair of gases. Although atomic structures of the studied pores remain unknown, type-3 pores are similar in size to hepta-vacancies (Supplementary Information) and match intrinsic pores in γ-graphyne. If/when 2D materials with such pores are developed, one can envisage separation technologies with unprecedented selectivities, beyond the existing selectivity-permeability bounds (for projections based on our results, see Fig. S5).

1. Wang, L. et al. *Nat. Nanotechnol.* **12**, 509–522 (2017).
2. Epsztein, R., DuChanois, R. M., Ritt, C. L., Noy, A. & Elimelech, M. *Nat. Nanotechnol.* **15**, 426–436 (2020).
3. Park, H. B., Kamcev, J., Robeson, L. M., Elimelech, M. & Freeman, B. D. *Science* **356**, eaab0530 (2017).
4. Koenig, S. P., Wang, L., Pellegrino, J. & Bunch, J. S. *Nat. Nanotechnol.* **7**, 728–732 (2012).
5. Wang, L. et al. *Nat. Nanotechnol.* **10**, 785–790 (2015).
6. O'Hern, S. C. et al. *Nano Lett.* **14**, 1234−1241 (2014).
7. Boutilier, M. S. H. et al. *ACS Nano* **11**, 5726–5736 (2017).
8. Zhao, J. et al. *Sci. Adv.* **5**, eaav1851 (2019).
9. Thiruraman, J. P. et al. *Sci. Adv.* **6**, eabc7927 (2020).
10. Qiu, H., Xue, M., Zhang, Z. & Guo, W. *Adv. Mater.* **31**, 1803772 (2019).
11. Gao, X., Liu, H., Wang, D. & Zhang, J. *Chem. Soc. Rev.* **48**, 908–936 (2019).
12. Neumann, C. et al. *ACS Nano* **13**, 7310−7322 (2019).
13. Robeson, L. M. *J. Membr. Sci.* **320**, 390–400 (2008).
14. Jiang, D.-E., Cooper, V. R. & Dai, S. *Nano Lett.* **9**, 4019–4024 (2009).
15. Blankenburg, S. et al. *Small* **6**, 2266–2271 (2010).
16. Celebi, K. et al. *Science* **344**, 289–292 (2014).
17. Sun, C. et al. *Langmuir* **30**, 675–682 (2014).
18. Yuan, Z. et al. *ACS Nano* **11**, 7974–7987 (2017).
19. Yuan, Z., Misra, R. P., Rajan, A. G., Strano, M. S. & Blankschtein, D. *ACS Nano* **13**, 11809–11824 (2019).
20. Vallejos-burgos, F., Coudert, F.-X. & Kaneko, K. *Nat. Commun.* **9**, 1812 (2018).
21. Sun, P. Z. et al. *Nature* **579**, 229–232 (2020).
22. Krasheninnikov, A. V. & Banhart, F. *Nat. Mater.* **6**, 723–733 (2007).
23. Yuzvinsky, T. D., Fennimore, A. M., Mickelson, W., Esquivias, C. & Zettl, A. *Appl. Phys. Lett.* **86**, 053109 (2005).
24. Sommer, B. et al. *Sci. Rep.* **5**, 7781 (2015).
25. Chen, X. et al. *Nano Lett.* **20**, 8, 5670–5677 (2020).
26. Schweizer, P. et al. *Nat. Commun.* **11**, 1743 (2020).
27. Mehio, N., Dai, S. & Jiang, D. E. *J. Phys. Chem. A* **118**, 1150–1154 (2014).
28. Freeman, B. D. *Macromolecules* **32**, 375–380 (1999).
29. Robeson, L. M., Freeman, B. D., Paul, D. R. & Rowe, B. W. *J. Membr. Sci*. **341**, 178–185 (2009).

## Supplementary Information

### 1. Device fabrication and inspection

To make our devices and test their atomically tight sealing, we followed the procedures developed in ref. 1. In brief, monocrystals of graphite with a thickness of >200 nm were prepared by mechanical exfoliation on an oxidized silicon wafer. The crystals were examined in an optical microscope using both dark-field and differential-interference-contrast modes to locate relatively large areas (over tens of microns in size), which were free from wrinkles, folds, atomic-step terraces or other defects. Then, using electron-beam lithography and dry etching, an array of microwells with internal diameters of 1 to 3 μm and depth of ~100 nm was fabricated within the found atomically flat areas. After overnight annealing at 400°C in $H_2$/Ar atmosphere (volume ratio of 1:10), the microwells were sealed with a large crystal of monolayer graphene, which was transferred in ambient air (Fig. 1). The resulting devices were carefully inspected using AFM, and those showing any damage to their sealing were discarded. Such damage could be, for example, extended defects in the atomically flat top surface of the microwells or wrinkles in the graphene sealing[1]. The remaining devices were leak-tested by placing them into a stainless-steel chamber containing Ar or Kr at a partial pressure $P \approx 3$ bar. After a few days, they were taken out and quickly (typically within 3 mins) checked using AFM for any changes in $\sigma$ (here $\sigma$ the central position of the curved membrane along the microwell's cylindrical axis $z$, and $z$ is assumed to be zero at cavity's top surface; see Fig. 1b). Again, we discarded those devices that exhibited any sign of leakage, namely, if changes in the membrane position $\Delta\sigma$ after pressurization were > 1 nm. Finally, we repeated the same leak test but in an atmosphere of helium at 1 bar. Only devices with no changes in membrane positions were kept for further investigation.

### 2. Perforating graphene with low-energy electrons

Devices that successfully passed the above inspection were exposed to electron irradiation in a scanning electron microscope (*Zeiss EVO MA10*). The accelerating voltage was chosen between 2 and 10 kV, and the beam current was set at ~10 pA. In a single exposure, the membranes were irradiated at a magnification of ×700 for 3–5 seconds, which translated into an electron dose of 0.1 – 0.2 $\mu C\ cm^{-2}$ or only ~$10^4$ electrons per $\mu m^2$. After the exposure, the devices were subjected to the same leak tests as described in section 1. We repeated this exposure-test cycle several times until the irradiated container started to exhibit a leak, indicating a defect created in its graphene membrane. Note however that, in about 20% of cases, we could not create any discernable leak, no matter how long the membranes were exposed to the electron beam. We believe this was due to crosslinking of hydrocarbons under electron irradiation, which protected the graphene surface from water-mediated damage. This is the same passivation mechanism[2,3] as discussed in the main text, but the hydrocarbon protection developed before a pore in graphene appeared.

### 3. Evaluation of permeation rates

The permeation rates $\Gamma$ were evaluated using the initial linear slopes of $\sigma(t)$ (see Fig. 1d) following the approach detailed in refs. 1,4,5. Briefly, the pressure inside the microcavities is the sum of the atmospheric pressure $P_a$ coming from the trapped air and the initial partial pressure $P(0)$ of the tested gas. The resulting differential pressure $P$ acting on a graphene membrane is given by Hencky's solution for a clamped circular membrane[6]

$$P(t) = \frac{\kappa Y L \sigma(t)^3}{a^4} \quad \text{(S1)}$$

where $\kappa \approx 3$ is a coefficient that depends on Poisson's ratio[4], $Y$ is Young's modulus, $L$ is the membrane thickness, and $a$ is the radius of the cavity with a depth $H$. According to the ideal gas law

$$(P_a + P)V = (n_{air} + n_{gas})N_A k_B T \quad \text{(S2)}$$

where $n_{air}$ and $n_{gas}$ are the number of moles of air and tested gas inside the container, respectively, and $N_A$ is the Avogadro constant. The total volume of the trapped gas $V = \pi a^2H + c\pi a^2\sigma$ is given by microwell's volume, $\pi a^2H$, and the additional volume $c\pi a^2\sigma$ due to the bulging membrane, where $c \approx 0.5$ is a geometric constant accounting for the membrane curvature. For simplicity, we neglect the volume's reduction caused by sagging and, for the purpose of this section only, count $\sigma$ from its unpressurized state ($P = 0$) rather than the top graphite surface.
By differentiating eq. S2 while utilizing eq. S1, we obtain

$$\Gamma \equiv -\frac{dn_{gas}}{dt} \approx -\frac{\pi a^2}{N_A k_B T}\frac{d}{dt}(P_a c\sigma + PH + Pc\sigma) = \frac{\pi a^2}{N_A k_B T}(P_a c + 3PH/\sigma + 4Pc)\left|\frac{d\sigma}{dt}\right| \quad \text{(S3)}$$

where we also used the fact that, because of the equal partial pressures of air outside and inside the container, there is little flow of air through the pore, $\frac{dn_{air}}{dt} \approx 0$. The flow rate $\Gamma$ on the left-hand side of eq. S3 is generally proportional to $P$ and, therefore, $\propto \sigma^3(t)$ as per eq. S1, whereas the second and third terms on the right-hand side are proportional to $\sigma^2$ and $\sigma^3$, respectively. Accordingly, eq. S3 is a nonlinear but autonomous differential equation. It allows an analytical solution $\sigma(t)$ but, because we are interested only in the initial response, it is straightforward to show that, initially, $\sigma$ evolves linearly with time, $\frac{d\sigma}{dt}(0)= const.$ The solution is given by eq. S3 at $t = 0$, that is,

$$\Gamma(0) = \frac{\pi a^2}{N_A k_B T}\left[P_a c + \frac{3P(0)H}{\sigma(0)} + 4P(0)c\right]\left|\frac{d\sigma}{dt}(0)\right| \quad \text{(S4)}$$

Generally, the linearity should hold only for small changes in $\sigma$, but our modelling using the full solution of eq. S3 shows that, if $P(0)$ is above a few bars, the fitting of the experimental curves with a linear dependence using eq. S4 should result in $\Gamma$ accurate within a factor of < 2, even if $\sigma$ changes by as much as 40%. Furthermore, for our usual case of $P(0) > P_a$ and $H > \sigma$, the first term in the brackets in eq. S4 can be neglected and, therefore, the permeation rates should depend linearly on $P$, which is consistent with our assumption above and agrees with the experiment (Fig. S1). This also allows us to introduce the permeance $\Gamma^* = \Gamma/P$, which characterizes the pores, independently of $P(0)$ used for pressurization.

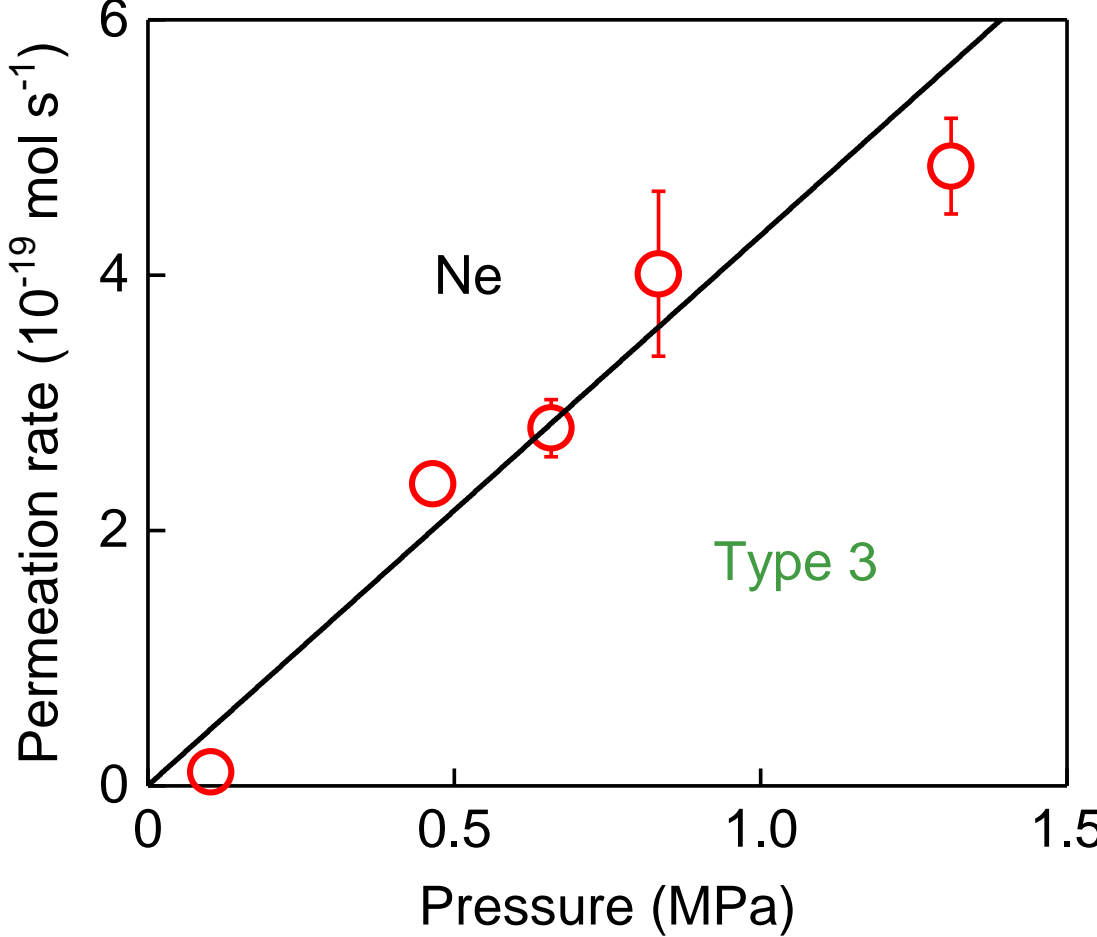

***Figure S1| Pressure dependence.*** *Permeation rates for Ne at different P using a type-3 pore. Symbols: experimental data. Solid curve: best linear fit. Error bars: SD for linear-in-time fitting of σ and are shown only if larger than symbols.*

## 4. Counting the pores

To estimate the number of pores that were created in our membranes by the electron beam, we followed the approach described in ref. 5. To this end, we sealed the pores with Au nanoparticles by depositing a small amount of Au on top of graphene membranes using thermal evaporation. In a single cycle, the evaporation lasted for 3 s at a rate of 0.3 Å $s^{-1}$, which resulted in Au nanoparticles covering < 1% of the membrane area, in agreement with ref. 5. Then the devices were pressurized with Ar, and $\sigma(t)$ was monitored by AFM. This procedure was repeated several times. As exemplified in Fig. S2, we typically observed a sharp decrease in $\Gamma^*$ after one or two deposition cycles (>95% of cases), and $\Gamma^*$ remained essentially constant after further cycles. This observation agrees with the expectations for only a single pore being present in the graphene membranes. Indeed, for multiple pores, $\Gamma^*$ should decrease not in a sudden jump but stepwise with increasing the density of Au particles because, at such a low areal coverage of <1 %, it would be practically impossible to block more than one pore in a single deposition cycle[5].

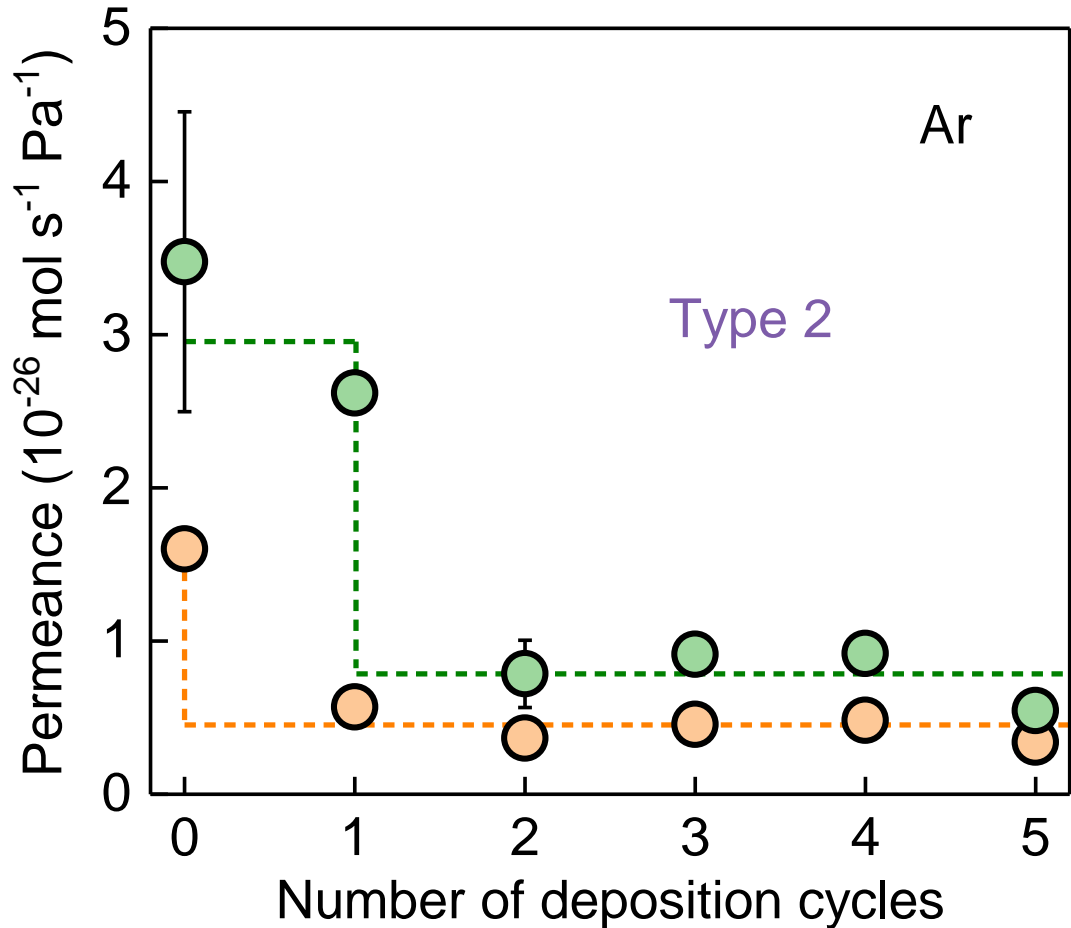

***Figure S2| Sealing the angstrom pores in graphene.*** *Permeance of Ar through type-2 pores during repeated deposition of Au nanoparticles (colors mark two different devices). The error bars are SD for linear fits of $\sigma(t)$. Dashed curves: guides to the eye. Note that Au nanoparticles placed on top of pores do not seal them completely, and $\Gamma^*$ is typically suppressed by a factor of ~4, close to the observations reported in ref. 5.*

## 5. Effect of high radiation doses

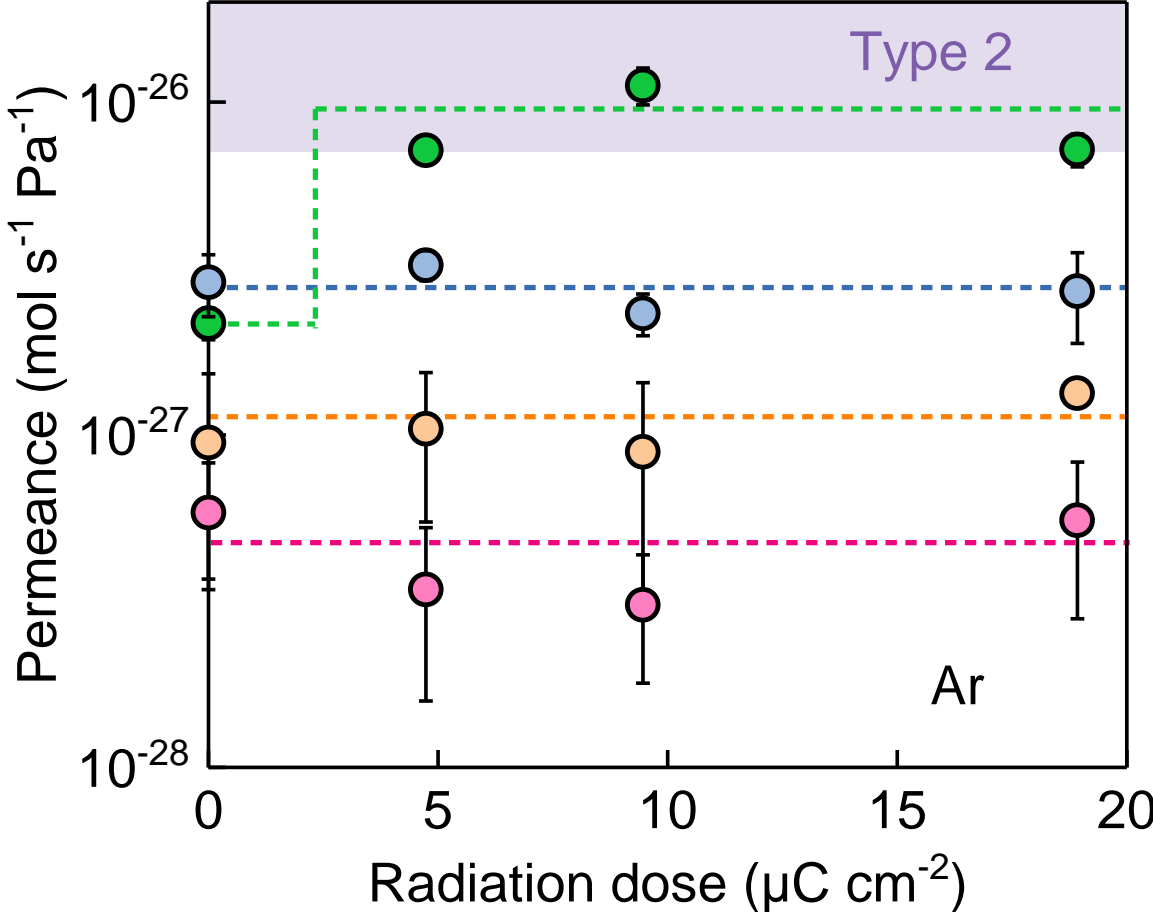

***Figure S3| Effect of additional irradiation.*** *Initially all the devices (different colors) had the smallest (type-3) pores created by doses of <0.5 μC $cm^{-2}$. Then the graphene membranes were subjected to further electron radiation (doses were up to 100 times higher). Less than 20% of pores exhibited discernable changes in their permeation rates, as exemplified by the device shown in green. Its pore seemed to evolve from type 3 to type 2. Symbols: experimental data with error bars indicating SD for the linear $\sigma(t)$ fits. Dashed lines: guides to the eye. The shaded area refers to the range of Ar permeances observed for type-2 pores.*

Several studied devices with pores identified as type 3 were later exposed to electron-beam radiation at 8 keV. Most of them (> 80%) exhibited stable $\Gamma^*$ even after radiation doses of 100 times larger (Fig. S3). This indicated that no additional pores appeared and no modification to the existing type-3 pores occurred. Only in a few occasions, we observed a notable increase in permeation rates, which fell into the range of $\Gamma^*$ typical for type-2 pores (green curves in Fig. S3). We speculate that such pores probably developed from the original type-3 pores (rather than extra pores were created) because the additional radiation had negligible effect in most cases whereas the existing pores could serve as weak spots allowing further electron damage.

### 6. Ab-initio simulations for gas translocation through graphene pores

Trying to evaluate the energy barrier $E$ for translocation of a noble atom through graphene pores, we performed first-principles calculations in the framework of density functional theory (DFT), as implemented in the Vienna *ab-initio* simulation package (VASP)[7]. The Perdew-Burke-Ernzerhof (PBE) form of generalized gradient approximation (GGA) was adopted to describe the electron exchange and correlation[8]. The van der Waals (vdW) correction to the GGA functional was included using the DFT-D2 method of Grimme.

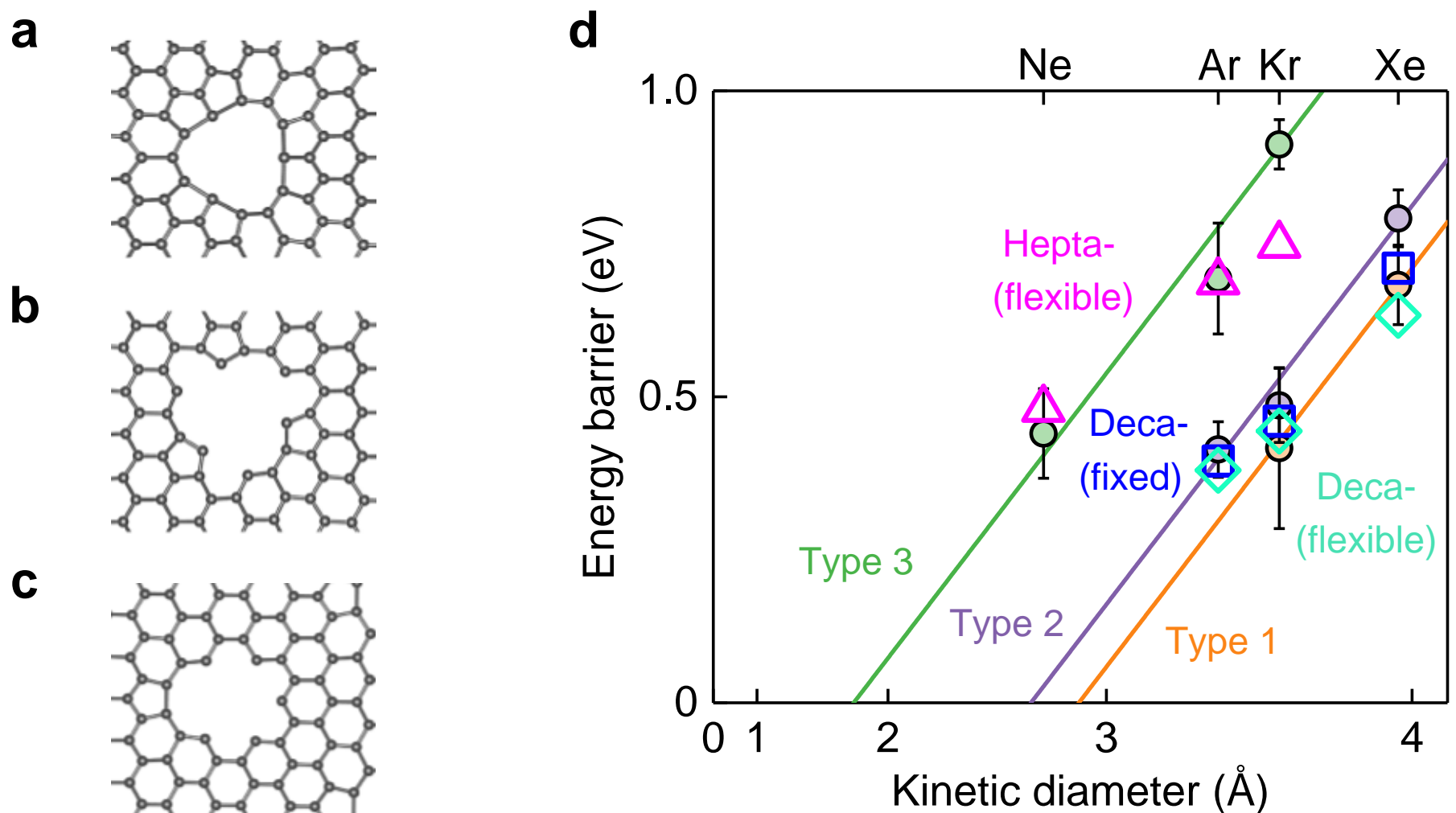

***Figure S4| Simulations of gas permeation through graphene pores. a, b,*** *Schematics for the pore formed by removing 10 carbon atoms in its initial (**a**) and reconstructed (**b**) configurations.* ***c****, Pore formed by removing 7 carbon atoms (shown is its reconstructed configuration).* ***d****, The barriers E for translocation of Ne, Ar, Kr and Xe through the reconstructed pores in (**b**, **c**). During translocation, carbon atoms of the deca-vacancy were either fixed (blue squares) or allowed to elastically deform (cyan diamonds). Pink triangles: E for deformable hepta-vacancy. The calculated E are compared with* $E_A$ *found experimentally (color-coded symbols and lines from Fig. 3b of the main text).*

To construct graphene pores of different sizes, a certain number of carbon atoms were removed from the graphene lattice followed by bond reconstruction to eliminate undercoordination of the edge atoms. To be specific, Stone Wales (SW or 57) defect, 585- and 555-777- divacancies, tetra- and hexa- vacancies were created based on a 5×5 graphene supercell, whereas the largest deca-vacancy was created in a 6×6 supercell. The pore area obviously increased with the number of removed carbon atoms. For example, the largest defect sites (pores) for SW and 555-777-divacancy were 7-membered rings, whereas those in 585-divacancy, tetra-, hexa- and deca-vacancies were 8-, 9-, 10- and 12-membered rings, respectively. As an example, the deca-vacancy is shown in Fig. S4a. The geometric areas $A_n$ of these pores as determined by the position of carbon atoms on the edges (where $n$ is the number of edge atoms) were $A_7 \approx 7.6$ and 7.9 Å$^2$ for SW and 555-777-divacancy, $A_8 \approx 11.0$ Å$^2$ for 585-divacancy, $A_9 \approx 14.3$ Å$^2$ for tetra-vacancy, $A_{10} \approx 17.1$ Å$^2$ for hexa-vacancy and $A_{12} \approx 28.0$ Å$^2$ for deca-vacancy.

However, because of the electronic clouds surrounding carbon atoms, the penetrating gas atoms should effectively experience considerably smaller openings in the graphene lattice. To disentangle graphene's electronic and elastic contributions to the energy barrier *E*, we first deliberately fixed the positions of carbon atoms around the pores (fixed configuration), which revealed the effect of electron-cloud distortion caused by squeezing gas atoms through. In the next step, we allowed the lattice to be fully flexible in all three directions (flexible configuration), which yielded the combined effect of electron-cloud and elastic lattice distortions. For details of our numerical simulations, we refer to the previous report[9].

We found that, in both fixed and flexible configurations, the barrier *E* for gas transport through all the studied pores scaled well with $d_K^2$, in agreement with our experiment. All non-reconstructed pores with small $A_n$ ($n$ = 7-12) yielded *E* considerably higher than $E_A$ found experimentally. For example, in its non-reconstructed state (Fig. S4a), the deca-vacancy provided $E \approx 1.5$, 2.2 and 3.5 eV for Ar, Kr and Xe atoms, respectively. As the next step, the vacancy configurations were allowed to fully relax in all three directions. We found that small pores remained approximately circular. In contrast, our largest pore (deca-vacancy) experienced notable edge reconstruction, as illustrated in Fig. S4b. This resulted in its larger pore size ($A \approx 46.8$ Å$^2$) and reduced stress. We found that *E* through this reconstructed pore were close to experimental $E_A$ for type-1 and type-2 pores, regardless of whether the positions of rim atoms were fixed or flexible during translocation (Fig. S4d). In comparison with the fixed pore in Fig. S4b, its flexible configuration led to a reduction in *E* of maximum 11% for the largest atom of Xe. This gives a general idea about the amount of elastic energy involved in translocation of gases through atomic-scale pores. Using the same approach, we also constructed an intermediate size pore by removing 7 carbon atoms and, after full relaxation, the resulting hepta-vacancy (Fig. S4c) had the geometric area $A \approx 38.7$ Å$^2$. In their flexible configuration, the calculated barriers *E* closely matched $E_A$ found for type-3 pores in our experiments (Fig. S4d).

### 7. Permeation dominated by surface-adsorbed gases

Permeating gas particles could either come from the bulk phase, striking directly at pore's area *A*, or diffuse into pore's mouth after first being adsorbed on graphene within a finite-size area $A_{fin}$ around the pore. Accordingly, the permeation rate can be expressed as $\Gamma = \Gamma_{bu} + \Gamma_{ad}$ where $\Gamma_{bu}$ and $\Gamma_{ad}$ denote the bulk and adsorbed-phase contributions, respectively. The former is described by[10]

$$\Gamma_{bu} = \frac{\upsilon_0}{N_A}\exp\left(-\frac{E}{k_B T}\right) = \frac{A}{N_A}\frac{P}{\sqrt{2\pi m k_B T}}\exp\left(-\frac{E}{k_B T}\right) \tag{S5}$$

where $\nu_0$ is the attempt rate from the bulk phase, that is, the number of gas atoms or molecules striking the pore area each second. Similarly, the permeation rate from the adsorption phase can be written as

$$\Gamma_{ad} = \frac{\upsilon_{ad}}{N_A}\exp\left(-\frac{E}{k_B T}\right) \tag{S6}$$

where $\nu_{ad}$ is the attempt rate for adsorbed atoms or molecules, which can in turn be expressed as[11]

$$\upsilon_{ad} = \sqrt{\frac{k_B T}{2\pi m}}\rho C \tag{S7}$$

where $\rho$ is the areal density of adsorbed gases and *C* is the pore circumference, $C \approx \pi d_P$. In this equation, it is assumed that absorbed atoms or molecules form an ideal 2D gas on the graphene surface owing to their little interaction with atomic corrugations[12]. The areal density can be expressed[10] as $\rho \approx \frac{P}{\sqrt{2\pi m k_B T}}\frac{1}{f_d}$ where the desorption frequency $f_d$ can be written as[13,14] $f_d = f_0 K$ with $f_0 = k_B T/h$ being the frequency of molecular vibration and *h* the Planck constant. The thermodynamic equilibrium constant *K* is given by the van 't Hoff equation[13,14]

$$K = \exp\left(\frac{\Delta S}{k_B}\right)\exp\left(-\frac{E_{ad}}{k_B T}\right) \tag{S8}$$

where $\Delta S$ is the entropy change during the permeation process and $E_{ad}$ is the adsorption energy (in this form, $E_{ad}$ is positive). Combining the above equations, we obtain

$$\Gamma = \frac{1}{N_A}(\upsilon_0 + \upsilon_{ad})\exp\left(-\frac{E}{k_B T}\right) = \frac{1}{N_A}\frac{P}{\sqrt{2\pi m k_B T}}\left(A + \sqrt{\frac{k_B T}{2\pi m}}\frac{C}{f_d}\right)\exp\left(-\frac{E}{k_B T}\right) \tag{S9}$$

which is consistent with the linear $P$ dependence observed experimentally. The measured $T$ dependences allowed us to evaluate both attempt rates $\nu$ and the activation energy $E_A$ (see the main text). The found $\nu$ (Fig. 3c) were many orders of magnitude larger than $\nu_0$, indicating that the bulk phase contributes little to the observed permeation. Accordingly, eq. S9 can be simplified and further extended as

$$\Gamma = \frac{\upsilon_{ad}}{N_A}\exp\left(-\frac{E}{k_BT}\right) = \frac{P d_P h}{2N_A m k_B T}\exp\left(-\frac{\Delta S}{k_B}\right)\exp\left(-\frac{E-E_{ad}}{k_BT}\right) \quad \text{(S10)}$$

According to eq. S10, the permeation barrier $E$ should be larger than the measured activation energy $E_A = E - E_{ad}$. Furthermore, adsorbed atoms or molecules diffusing along the graphene surface and permeating through angstrom-scale pores are expected[15-17] to exhibit a considerable entropy loss $\Delta S < 0$ at the translocation position, which results in an enhancement of $\Gamma$ by a factor of $\exp(|\Delta S|/k_B)$, similar to the case of polymeric membranes[18,19].

## 8. Comparison with other gas-selective membranes

To compare potential performance of angstroporous 2D materials with that of membranes made from 3D materials, let us assume a pore density of $10^{14}$ cm$^{-2}$ or one pore per nm$^2$. Such a density is offered by several graphene allotropes[20-22]. It might also be possible (albeit difficult) to achieve comparable densities by defecting graphene (e.g., sub-nm pores with densities of $\sim 10^{12}$ cm$^{-2}$ were demonstrated in graphene using ion bombardment[23]). To allow comparison between 2D and 3D cases, we also assume for simplicity that all 3D membranes can be made down to ~100 nm in thickness, even though this is impossible for many of the materials without loss of functionality[24-26]. The results of our analysis are summarized in Fig. S5 that plots the experimentally observed selectivities as a function of permeance for 11 different pairs of gases (namely, $O_2/N_2$, $H_2/N_2$, $He/N_2$, $He/H_2$, $CO_2/N_2$, $H_2/CO_2$, $He/CO_2$, $CO_2/CH_4$, $H_2/CH_4$, $He/CH_4$ and $N_2/CH_4$). The literature data include membranes made from polymers[27-44], metal-organic frameworks[45-55], graphene oxide laminates[56-65], covalent organic frameworks[66-69], zeolites[70-78], transition metal dichalcogenides[79-83], MXenes[84,85], layered double hydroxides[86,87], carbon nitride[88], silica[89-92], silicon carbide[93] and carbon molecular sieves[94-97]. The figure also shows the results previously reported for porous graphene[98-103] and the current Robeson bounds for polymeric membranes[104].

One can clearly see in Fig. S5 that, in principle, the discussed angstroporous 2D materials promise performance that compares favorably with all the other membranes for gas separation. Particularly, angstroporous membranes can provide orders-of-magnitude improvements for separation and removal of relatively large gas molecules (such as $CH_4$ in the case of Fig. S5 and, also, Xe). This ability is particularly prominent for our smallest, type-3 pores that essentially block these molecules. Even at densities of $\sim 10^{10}$–$10^{12}$ cm$^{-2}$, which should be achievable by top-down approaches, graphene with type-3 or similar-size pores would offer superior performance, owing to the exponentially-high selectivity for gas molecules with $d_K$ differing by a factor of ~2. The sieving effect is less pronounced for smaller molecules, which shows that angstroporous 2D membranes offer notable advantages mostly in the activation regime ($d_P < d_K$). To achieve better selectivity-permeability tradeoff with respect to smaller molecules ($d_K < 3.5$ Å such as $O_2$ and $CO_2$ in Fig. S5), pores smaller than type-3 are required ($d_P \leq 1.5$ Å). Those are potentially available in some graphynes[20-22].

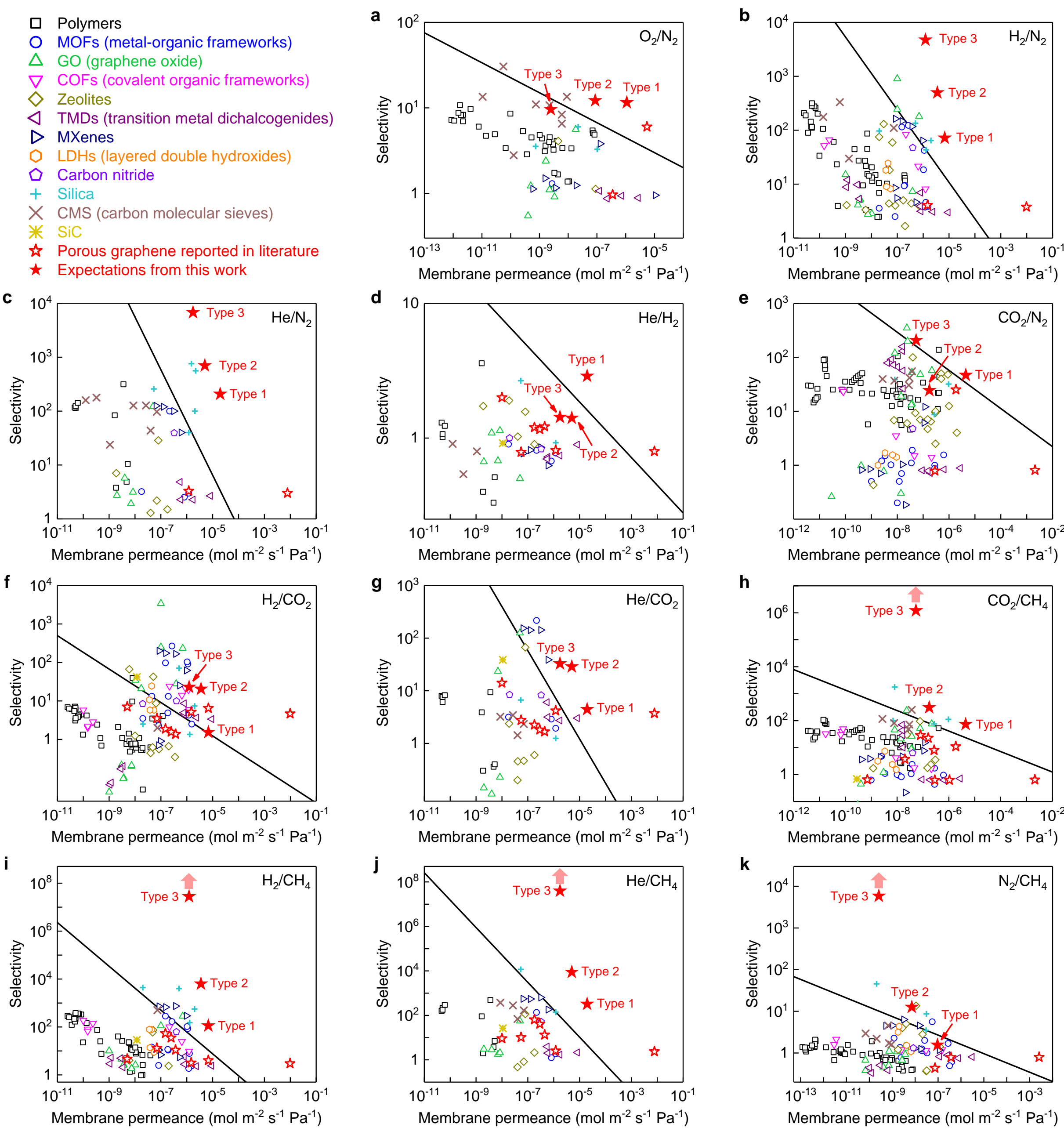

***Figure S5| Projected performance of angstroporous 2D materials.*** *(**a**-**k**) Selectivity for different pairs of gases as stated in the upper right corners of each panel. The solid-star symbols are for 2D membranes that are assumed to have type 1, 2 or 3 pores with densities of $10^{14}$ $cm^{-2}$. Empty stars: projected performances reported previously for porous graphene membranes. The other symbols are literature data as specified in the figure (top left). The black lines are the current Robeson bounds for polymers assuming their 100 nm thickness (adapted from ref. 104). The arrows in (**h**-**k**) indicate that we can provide only the minimal bounds for type-3 pores because of the limits on permeation of $CH_4$ and Xe through these pores.*

## Supplementary References

1. Sun, P. Z. et al. *Nature* **579**, 229–232 (2020).
2. Yuzvinsky, T. D., Fennimore, A. M., Mickelson, W., Esquivias, C. & Zettl, A. *Appl. Phys. Lett.* **86**, 053109 (2005).
3. Sommer, B. et al. *Sci. Rep.* **5**, 7781 (2015).
4. Koenig, S. P., Wang, L., Pellegrino, J. & Bunch, J. S. *Nat. Nanotechnol.* **7**, 728–732 (2012).
5. Wang, L. et al. *Nat. Nanotechnol.* **10**, 785–790 (2015).
6. Hencky, H. *Z. Math. Phys.* **63**, 311–317 (1915).
7. Gresse, K. & Furthmuller, F. *Phys. Rev. B* **54**, 11169 (1996).
8. Perdew, J. P., Burke, K. & Ernzerhof, M. *Phys. Rev. Lett.* **77**, 3865–3868 (1996).
9. Griffin, E. et al. *ACS Nano* **14**, 7280–7286 (2020).
10. Landau, L. D. & Lifshitz, E. M. Course of theoretical physics. Volume 5 (1980).
11. Yuan, Z., Misra, R. P., Rajan, A. G., Strano, M. S. & Blankschtein, D. *ACS Nano* **13**, 11809–11824 (2019).
12. Bartolomei, M. et al. *J. Phys. Chem. C* **117**, 10512–10522 (2013).
13. Eyring, H. *J. Chem. Phys.* **3**, 107–115 (1935).
14. Hanggi, P., Talkner, P. & Borkovec, M. *Rev. Mod. Phys.* **62**, 251–342 (1990).
15. Campbell, C. T. & Sellers, J. R. *J. Am. Chem. Soc.* **134**, 18109–18115 (2012).
16. Dauenhauer, P. J. & Abdelrahman, O. A. *ACS Cent. Sci.* **4**, 1235–1243 (2018).
17. Meares, P. *J. Am. Chem. Soc.* **76**, 13, 3415–3422 (1954).
18. Freeman, B. D. *Macromolecules* **32**, 375–380 (1999).
19. Robeson, L. M., Freeman, B. D., Paul, D. R. & Rowe, B. W. *J. Membr. Sci*. **341**, 178–185 (2009).
20. Qiu, H., Xue, M., Zhang, Z. & Guo, W. *Adv. Mater.* **31**, 1803772 (2019).
21. Gao, X., Liu, H., Wang, D. & Zhang, J. *Chem. Soc. Rev.* **48**, 908–936 (2019).
22. Neumann, C. et al. *ACS Nano* **13**, 7310−7322 (2019).
23. O'Hern, S. C. et al. *Nano Lett.* **14**, 1234−1241 (2014).
24. Wang, L. et al. *Nat. Nanotechnol.* **12**, 509–522 (2017).
25. Epsztein, R., DuChanois, R. M., Ritt, C. L., Noy, A. & Elimelech, M. *Nat. Nanotechnol.* **15**, 426–436 (2020).
26. Park, H. B., Kamcev, J., Robeson, L. M., Elimelech, M. & Freeman, B. D. *Science* **356**, eaab0530 (2017).
27. Li, S., Wang, Z., Yu, X., Wang, J. & Wang, S. *Adv. Mater.* **24**, 3196–3200 (2012).
28. Shen, Y., Wang, H., Liu, J. & Zhang, Y. *ACS Sustain. Chem. Eng.* **3**, 1819–1829 (2015).
29. Fu, Q. et al. *Energy Environ. Sci.* **9**, 434–440 (2016).
30. Du, N. et al. *Nat. Mater.* **10**, 372–375 (2011).
31. Park, H. B. et al. *Science* **318**, 254–258 (2007).
32. Shan, M. et al. *Sci. Adv.* **4**, 1698–1705 (2018).
33. Carta, M. et al. *Science* **339**, 303–307 (2013).
34. Rezac, M. E. & Schöberl, B. *J. Memb. Sci.* **156**, 211–222 (1999)
35. Nagai, K., Higuchi, A. & Nakagawa, T. *J. Polym. Sci., B, Polym. Phys.* **33**, 289–298 (1995).
36. Hamid, M. A., Chung, Y. T., Rohani, R. & Junaidi, M. U. *Sep. Purif. Technol.* **209**, 598–607 (2019).
37. Han, J. Y., Lee, W. S., Choi, J. M., Patel, R. & Min, B. R. *J. Membr. Sci.* **351**, 141–148 (2010).
38. Asghar, H., Ilyas, A., Tahir, Z., Li, X. & Khan, A. L. *Sep. Purif. Technol.*, 203, 233–241 (2018).
39. Giel, V. et al. *Eur. Polym. J.* **77**, 98–113 (2016).
40. Yong, W. F., Li, F. Y., Chung, T. S. & Tong, Y. W. *J. Membr. Sci.* **462**, 119–130 (2014).
41. Zhao, S., Liao, J., Li, D., Wang, X. & Li, N. *J. Membr. Sci.* **566**, 77–86 (2018).
42. Choi, S.-H., Tasselli, F., Jansen, J. C., Barbieri, G. & Drioli, E. *Eur. Polym. J.* **46**, 1713–1725 (2010).
43. Ghanem, B. S., McKeown, N. B., Budd, P. M. & Fritsch, D. *Macromolecules* **41**, 1640–1646 (2008).
44. Kosuri, M. R. & Koros, W. J. *J. Membr. Sci.* **320**, 65–72 (2008).
45. Peng, Y. et al. *Science* **346**, 1356–1359 (2014).
46. Li, Y. J., Liu, H., Wang, H. T., Qiu, J. S. & Zhang, X. F. *Chem. Sci.* **9**, 4132–4141 (2018).
47. Wang, X. et al. *Nat. Commun.* **8**, 14460 (2017).
48. Peng, Y., Li, Y., Ban, Y. & Yang, W. *Angew. Chem. Int. Ed.* **56**, 9757–9761 (2017).
49. Zhou, S. et al. *Int. J. Hydrogen Energy* **38**, 5338–5347 (2013).
50. Guo, H., Zhu, G., Hewitt, I. J. & Qiu, S. *J. Am. Chem. Soc.* **131**, 1646–1647 (2009).
51. Aguado, S. et al. *New J. Chem.* **35**, 41–44 (2011).

52. Zhang, X. et al. *Chem. Mater.* **26**, 1975–1981 (2014).
53. Li, Y., Liang, F., Bux, H., Yang, W. & Caro, J. *J. Membr. Sci.* **354**, 48–54 (2010).
54. Yoo, Y., Varela-Guerrero, V. & Jeong, H.-K. *Langmuir* **27**, 2652–2657 (2011).
55. Ranjan, R. & Tsapatsis, M. *Chem. Mater.* **21**, 4920–4924 (2009).
56. Shen, J. et al. *ACS Nano* **10**, 3398–3409 (2016).
57. Li, X. et al. *ACS Appl. Mater. Interfaces* **7**, 5528–5537 (2015).
58. Kim, H. W. et al. *Science* **342**, 91–95 (2013).
59. Li, H. et al. *Science* **342**, 95–98 (2013).
60. Zhou, F. et al. *Nat. Commun.* **8**, 2107 (2017).
61. Yang, J. et al. *Adv. Mater.* **30**, 1705775 (2018).
62. Cheng, L., Guan, K., Liu, G. & Jin, W. *J. Membr. Sci.* **595**, 117568 (2020).
63. Kim, H. W. et al. *Chem. Commun.* **50**, 13563–13566 (2014).
64. Ying, W. et al. *ACS Nano* **12**, 5385–5393 (2018).
65. Wang, S. et al. *Energy Environ. Sci.* **9**, 3107–3112 (2016).
66. Biswal, B. P., Chaudhari, H. D., Banerjee, R. & Kharul, U. K. *Chem. Eur. J.* **22**, 4695–4699 (2016).
67. Kang, Z. et al. *Chem. Mater.* **28**, 1277–1285 (2016).
68. Fan, H. et al. *J. Am. Chem. Soc.* **140**, 10094–10098 (2018).
69. Ying, Y. et al. *J. Mater. Chem. A* **4**, 13444–13449 (2016).
70. Varoon, K. et al. *Science* **334**, 72–75 (2011).
71. Tang, Z., Dong, J. & Nenoff, T. M. *Langmuir* **25**, 4848–4852 (2009).
72. Hong, M., Falconer, J. L. & Noble, R. D. *Ind. Eng. Chem. Res.* **44**, 4035–4041 (2005).
73. Kanezashi, M., O'Brien-Abraham, J., Lin, Y. S. & Suzuki, K. *AIChE J.* **54**, 1478–1486 (2008).
74. Kusakabe, K., Yoneshige, S., Murata, A. & Morooka, S. *J. Membr. Sci.* **116**, 39–46 (1996).
75. Shekhawat, D., Luebke, D. R. & Pennline, H. W. *US department of energy* (2003).
76. Kusakabe, K., Kuroda, T., Uchino, K., Hasegawaand, Y. & Mooroka, S. *AIChE J.* **14**, 1220–1226 (1999).
77. Poshusta, J. C., Tuan, V. A., Pape, E. A., Noble, R. D. & Falconer, J. L. *AIChE J.* **46**, 779–789 (2000).
78. Yu, M., Funke, H. H., Noble, R. D. & Falconer, J. L. *J. Am. Chem. Soc.* **133**, 1748–1750 (2011).
79. Wang, D., Wang, Z., Wang, L., Hu, L. & Jin, J. *Nanoscale* **7**, 17649–17652 (2015).
80. Shen, Y., Wang, H., Zhang, X. & Zhang, Y. *ACS Appl. Mater. Interfaces* **8**, 23371–23378 (2016).
81. Achari, A., Sahana, S. & Eswaramoorthy, M. *Energy Environ. Sci.* **9**, 1224–1228 (2016).
82. Chen, D., Ying, W., Guo, Y., Ying, Y. & Peng, X. *ACS Appl. Mater. Interfaces* **9**, 44251–44257 (2017).
83. Chen, D. et al. *J. Mater. Chem. A* **6**, 16566–16573 (2018).
84. Ding, L. et al. *Nat. Commun.* **9**, 155 (2018).
85. Shen, J. et al. *Adv. Funct. Mater.* **28**, 1801511 (2018).
86. Liu, Y., Wang, N., Cao, Z. & Caro, J. *J. Mater. Chem. A* **2**, 1235–1238 (2014).
87. Liu, Y., Wang, N. & Caro, J. *J. Mater. Chem. A* **2**, 5716–5723 (2014).
88. Villalobos, L. F. et al. *Sci. Adv.* **6**, eaay9851 (2020).
89. de Vos, R. M. & Verweij, H. *Science* **279**, 1710–1711 (1998).
90. Shelekhin, A. B., Dixon, A. G. & Ma, Y. H. *J. Membr. Sci.* **15**, 233–244 (1992).
91. Asaeda, M. & Yamasaki, S. *Sep. Purif. Technol.* **25**, 151–159 (2001).
92. Peters, T. A. et al. *J. Membr. Sci.* **248**, 73–80 (2005).
93. Elyassi, B., Sahimi, M. & Tsotsis, T. T. *J. Membr. Sci.* **288**, 290–297 (2007).
94. Shiflett, M. B. & Foley, H. C. *Science* **285**, 1902–1905 (1999).
95. Jones, C. W. & Koros, W. J. *Carbon* **32**, 1419–1425 (1994).
96. Yamamoto, M., Kusakabe, K., Hayashi, J. & Morooka, S. *J. Membr. Sci.* **133**, 195–205 (1997).
97. Centeno, T. A. & Fuertes, A. B. *Sep. Purif. Tech.* **25**, 379–384 (2001).
98. Boutilier, M. S. H. et al. *ACS Nano* **11**, 5726–5736 (2017).
99. Zhao, J. et al. *Sci. Adv.* **5**, eaav1851 (2019).
100. Celebi, K. et al. *Science* **344**, 289–292 (2014).
101. Choi, K., Droudian, A., Wyss, R. M., Schlichting, K.-P. & Park, H. G. *Sci. Adv.* **4**, eaau0476 (2018).
102. Huang, S. et al. *Nat. Commun.* **9**, 2632 (2018).
103. He, G. et al. *Energy Environ. Sci.* **12**, 3305–3312 (2019).
104. Robeson, L. M. *J. Membr. Sci.* **320**, 390–400 (2008).